# A ferroelectric memristor

André Chanthbouala[1], Vincent Garcia[1], Ryan O. Cherifi[1], Karim Bouzehouane[1], Stéphane Fusil[1,2], Xavier Moya[3], Stéphane Xavier[4], Hiroyuki Yamada[1,5], Cyrile Deranlot[1], Neil D. Mathur[3], Manuel Bibes[1], Agnès Barthélémy[1]* and Julie Grollier[1]

[1]Unité Mixte de Physique CNRS/Thales, 1 Av. A. Fresnel, Campus de l'Ecole Polytechnique, 91767 Palaiseau (France) and Université Paris-Sud, 91405 Orsay (France)

[2]Université d'Evry-Val d'Essonne, Bd. F. Mitterrand, 91025 Evry cedex (France)

[3]Department of Materials Science, University of Cambridge, Cambridge, CB2 3QZ (United Kingdom)

[4]Thales Research & Technology, 1 Av. A. Fresnel, Campus de l'Ecole Polytechnique, 91767 Palaiseau (France)

[5]National Institute of Advanced Industrial Science and Technology (AIST), Tsukuba, Ibaraki 305-8562 ( Japan)

*corresponding author : agnes.barthelemy@thalesgroup.com

**Memristors are continuously tunable resistors that emulate synapses[1,2]. Conceptualized in the 1970s, they traditionally operate by voltage-induced displacements of matter, but the mechanism remains controversial[3,4,5]. Purely electronic memristors have recently emerged[6,7] based on well-established physical phenomena with albeit modest resistance changes. Here we demonstrate that voltage-controlled domain configurations in ferroelectric tunnel barriers[8,9,10] yield memristive behaviour with resistance variations exceeding two orders of magnitude and a 10 ns operation speed. Using models of ferroelectric-domain nucleation and growth[11,12] we explain the quasi-continuous resistance variations and derive a simple analytical expression for the memristive effect. Our results suggest new opportunities for ferroelectrics as the hardware basis of future neuromorphic computational architectures.**

In tunnel junctions with a ferroelectric barrier, switching the ferroelectric polarization induces variations of the tunnel resistance, with resistance contrasts between the ON and OFF states of several orders of magnitude[9,13,14,15], defining a giant tunnel electroresistance (TER) effect. Several mechanisms have been proposed to explain this behaviour[16,17,18] but the dominant one appears related to changes in the tunnel barrier potential profile due to asymmetric polarization screening at barrier/electrode interfaces. In analogy with the operation of ferroelectric random access memories (FeRAMs), the large OFF/ON ratio in ferroelectric tunnel junctions (FTJs) has only been considered so far for binary data storage, with the key advantage of non-destructive readout and simpler device architecture. An important degree of freedom that has not yet been exploited in FTJs is the domain structure of the ferroelectric tunnel barrier. In ferroelectrics the domain size scales down with the square root of the film thickness[19,20], so that nanometre-size domains are expected for ferroelectric tunnel barriers (that are typically thinner than 5 nm). This provides a very fine level of control for the relative proportion of up and down domains and thereby of properties depending on the switched polarization.

Here we show that the domain configuration of a ferroelectric tunnel barrier can be controllably used to produce a virtually continuous range of resistance levels between OFF and ON states. We report piezoresponse force microscopy (PFM) images and electrical transport measurements as a function of the amplitude, duration and repetition number of voltage pulses in the 10-200 ns range. In a simple picture of conduction in parallel by up and down domains, we argue that the resistance variations are ruled by ferroelectric domain dynamics during polarization reversal. We analyse both OFF to ON and ON to OFF switching processes and model them in terms of domain nucleation and propagation. We conclude that FTJs emerge as a novel class of memristive systems for which state equations can be derived from models of polarization dynamics.

Our FTJs are composed of BaTiO$_3$(2 nm)/La$_{0.67}$Sr$_{0.33}$MnO$_3$(30 nm)/ (BTO/LSMO) extended layers on which Co/Au pads are defined by electron beam lithography (typical diameter 350 nm), sputtering and lift-off. Details on the growth and fabrication methods, as well as the demonstration of the ferroelectric properties of the BTO barrier have been given elsewhere [9,14]. Electrical contact to the pads was made using an AFM conductive tip. The measurements are performed by applying short (t$_{pulse}$ = 10-200 ns) write voltage pulses (of amplitude V$_{write}$) between the tip and the bottom electrode and subsequently measuring the tunnel resistance at low dc voltage (|V$_{read}$| = 100 mV).

In Figure 1a, we plot the junction resistance as we vary the amplitude of the applied voltage pulses while keeping a fixed pulse duration of 20 ns. A hysteretic cycle between low (R$_{ON}$ ~1.6 10$^5$ Ω) and high (R$_{OFF}$ ~ 4.6 10$^7$ Ω) resistance states is observed, with a large OFF/ON ratio of ~ 300 when the write voltage is swept between +4.2 and -5.6 V (Figure 1a, blue curve). Following previous results[14], the low resistance state (R$_{ON}$) corresponds to the ferroelectric polarization pointing up (P$_↑$), *i.e.* towards the Co/Au pad, which is also the virgin state for all devices. The switching between the two states is bipolar and, interestingly, not abrupt *i.e.*, a broad range of intermediate resistance states are observed. An asymmetry in the switching is visible and may reflect the presence of downward-polarized interfacial dipoles that favours the initial growth of downward polarized domains[21]. The minor loops in Figure 1a (cyan to red curves) show that depending on the cycling protocol the final resistance state can be finely tuned between R$_{ON}$ and R$_{OFF}$.

To get insight into the microscopic mechanisms responsible for this memristive effect, we have collected PFM images after poling the junctions into different resistance states, see Fig. 1b. Starting from a virtually homogeneous up-polarized state corresponding to a low resistance value (R= 3.10$^5$ Ω, i.e. close to the ON state , state ❶) , the application of positive voltage pulses nucleates down-polarized domains (white contrast in the red-framed images, state ❷ to ❺). Applying consecutive pulses of increasing amplitude causes the expansion of these existing down-polarized domains as well as the nucleation of new ones, consistent with previous results on thick ferroelectric films[22,23]. An almost saturated down state (R= $2.10^7$ Ω, i.e. close to the OFF state) is eventually reached (state ❺). Reversibly, applying negative pulses leads to a decrease in the resistance and to the correlated nucleation and propagation of up domains (dark contrast in the blue-framed images, states ❻ to ❽).

Overall, the junction resistance shows a systematic variation with the relative fraction of down domains extracted from the PFM images (red and blue symbols in Fig. 1b). This variation can be well reproduced in a simple model (black curve in Fig. 1b) considering that up- and down-polarized regions with different specific resistance conduct current in parallel (see the sketch in the inset of Fig. 1b). Thus, in ferroelectric tunnel junctions, a memristive behaviour can be devised by controlling the nucleation and growth of ferroelectric domains.

The junction resistance not only depends on the pulses amplitude but also on their duration and repetition number. Figure 2 presents phase diagrams of the resistance *vs.* pulse duration (in the 10-200 ns range) and pulse number shown in Figure 2 for three different pulse amplitudes. Along with the bipolar switching behaviour, this possibility to vary the resistance through the application of pulse sequences enables a simple scheme to continuously decrease or increase the junction resistance. We applied consecutive trains of positive and negative pulses (Fig. 3) after poling the junction in the ON state (reset). In Figure 3a-b, we fixed the number of consecutive positive pulses to 10 (amplitude : +2.9 V), which reproducibly set the junction into an intermediate resistance level of $4.10^6$ Ω. Clearly, the number of negative pulses (-2.7 V) applied subsequently determines the resistance level of the final state. In Figure 3c-d, we varied the number of consecutive positive pulses (+3 V) and fixed the number of negative ones (-3 V). The resistance gradually increases with the number of positive pulses, confirming the cumulative effects seen in Figure 2. Besides, the resistance after each negative pulses sequence depends on the level reached after the previous positive pulse train. Resistance switching from OFF to ON always appears more abrupt than from ON to OFF, which correlates with the asymmetry of the resistance *vs.* voltage cycle in Figure 1a.

Overall, the results shown in Fig. 3 demonstrate that the resistance level of the FTJ can not only be set by one pulse of appropriate amplitude but also by an appropriate number of consecutive pulses of a fixed voltage. This matches the definition of memristive devices and this latter functionality is particularly appealing for the integration of FTJs in brain-inspired computational architectures. In the particular case of artificial synapses, this would allow the modification of synaptic transmission through spike timing dependent plasticity, *i.e.* depending on the respective timing of spikes emitted by the pre and post neurons[4,24].

We now analyse the dynamics of resistance switching from ON to OFF and from OFF to ON. Assuming conduction in parallel for regions with $P_\downarrow$ or $P_\uparrow$ (Fig. 1e), we define the relative fraction of down domains by $s = (1/R - 1/R_{ON})/(1/R_{OFF} - 1/R_{ON})$; thus, $s$ varies from 0 in the ON state ($P_\uparrow$) to 1 in the OFF state ($P_\downarrow$). Figure 4 shows a typical set of data on the evolution of $s$ as a function of cumulative pulse time for pulse durations of 10 ns. For positive (negative) pulses, the initial state was initialized to $R_{ON}$ ($R_{OFF}$) corresponding to $P_\uparrow$ ($P_\downarrow$). While the polarization reversal starts immediately after the first pulse for up-to-down switching (Fig. 4d-f), it is delayed in the down-to-up case with a delay time that depends on the applied voltage (Fig. 4a-c). For both switching directions, $s$ does not always evolve smoothly toward the final state but presents a more "wavy" dependence. This signals the presence of several areas with different switching dynamics. This could be due to the sub-micron lithographic process we use to define FTJs, that may introduce a slight polarisation disorder, consistent with Gruverman *et al.*[23].

Two main models have been developed to describe the physics of ferroelectric polarization reversal by nucleation and propagation of domain walls. The Kolmogorov-Avrami-Ishibashi (KAI) model[11,12] applies to systems where switching is mainly driven by propagation, which is typically the case for clean epitaxial systems[25]. On the contrary, the nucleation-limited-switching models[26,27] have been developed to describe the dynamics of systems where switching is dominated by nucleation effects in disordered systems[28]. The delayed onset of switching at negative voltage can be ascribed to asymmetric nucleation processes: for up-to-down switching, pinned domains with down polarization serve as pre-existing nucleation centres; on the contrary, for down-to-up switching, nucleation centres need to be activated, explaining the observed delays in the *s vs.* time data (Figs. 4a-c) corresponding to increased nucleation times[21]. We argue that the remainder of the switching process occurs in the propagation regime. This picture discards an interpretation in terms of a purely nucleation-limited scenario.

To account for the observed "wavy" behaviour we model the data by dividing the pad area in a finite number of zones with different propagation and nucleation kinetics (different domain wall propagation speed, nucleation time, number of nuclei), each ruled by the KAI model. For a given zone i, we suppose that all nucleation sites are activated at the same nucleation time $\tau_N^i$ and then propagate with a characteristic propagation time $\tau_P^i$ that both depend on the voltage. Following this set of assumptions, the fraction of switched domains *s* can be written as

$$s = \sum_{i=1}^{N(i)} S_i * h(t - \tau_N^i) * \left\{ 1 - \exp\left[ -\left( \frac{t - \tau_N^i}{\tau_P^i} \right)^2 \right] \right\} \quad (1)$$

for up-to-down switching and as

$$s = 1 - \sum_{i=1}^{N(i)} S_i * h(t - \tau_N^i) * \left\{ 1 - \exp\left[ -\left( \frac{t - \tau_N^i}{\tau_P^i} \right)^2 \right] \right\} \quad (2)$$

for down-to-up switching, with $S_i$, the area of each zone normalized by the total junction area (with $\sum_{i=1}^{N(i)} S_i = 1$), and $h(t)$ the Heaviside step function.

Figure 4a, b, c (d, e, f) also show the fit of the experimental data by Eq. (2) [Eq. (1)] for negative (respectively positive) applied voltage with amplitude 2.25, 2.5 and 2.75 V. The data are well fitted on the whole time range; in particular the "wavy" dependence of *s vs.* time is accurately reproduced with a reduced number of zones $N \leq 5$. From the fits, we can extract for each zone the nucleation and propagation times that are plotted as a function of electric field for negative (Figs. 4g and 4h) and positive bias (Figs. 4i and 4j). The size of each symbol is proportional to the area $S_i$ of the zone it represents.

We first focus on the case of negative bias (Figs. 4g and 4h). For a given applied voltage, the same propagation time can be ascribed to all zones, whereas the nucleation times are scattered. Each zone is thus characterized by its nucleation properties, that can differ due to different densities and/or activation energies of nucleation centres. Nevertheless, after nucleation, the propagation is homogeneous over the surface of the junction. Both nucleation and domain wall propagation follow Merz's law, being proportional to $\exp(-E_a/E)$ (Figs. 4g and 4h), in agreement with other works[29,30]. The activation fields for nucleation $E_a(N)$ and propagation $E_a(P)$ are of the same order of magnitude, respectively 2.6 $10^{10}$ and 2.0 $10^{10}$ V/m. Reported values for the activation fields are quite dispersed in literature, ranging from 4 $10^5$ V/m in bulk to about 2 $10^8$ V/m for nanocapacitors with dimensions similar to our pads, but with a film thickness of 35 nm [Ref. 29]. The larger values we find here may be due to the highly strained character of our ultrathin ferroelectric films, in agreement with the large measured coercive fields[14,31].

For positive bias, the switching process is very different. As can be seen from the curves of Fig. 4 d, e and f, most of the switching occurs at very short time scales. The majority of nucleation events occurs within the first ten nanoseconds, below the resolution limit of our set-up. They correspond on Fig. 4i to the symbols (arbitrarily) positioned at t = 1 ns. These very small nucleation times are consistent with the already-mentioned possible presence of pinned domains with down polarization acting as pre-existing nucleation sites[21]. These early nucleated zones expand very fast, as can be seen from the corresponding short propagation times below 30 ns (blue squares in Fig. 4j). Different scenarios can account for this fast propagation, for example a very large density of the pre-existing nucleation centres, and/or an additional contribution to the total electric field coming from the pinned domains with down polarization.

Although most of the junction area is switched through this fast nucleation and propagation process, the remainder is reversed with dynamics that are globally comparable to the one observed for negative bias. Indeed, as observed in Fig. 4i, for a few zones representing a small fraction of the total surface the nucleation by positive voltage is delayed well beyond 10 ns. For these latecomers that did not expand from pre-existing nucleation centres, the nucleation and propagation times are quite similar to the ones measured with negative applied voltages (the small symbols in Figs. 4i and 4j fall close to dotted line corresponding to the negative bias fits).

Very importantly, the quantitative agreement between our model and the experimental data provides us with an analytical expression accounting for the observed memristive response based on the physical description of the ferroelectric domain dynamics. The memory effect typical of memristive systems can be defined by the following two equations[32]:

$$V(t) = \mathcal{R}(\sigma, V, i)\, i(t) \qquad (1)$$

$$d\sigma/dt = f(\sigma, V, t) \qquad (4)$$

where $\sigma$ represents one or several state variables, V is the voltage, i the current, t the time and $\mathcal{R}$ and $f$ are system-dependent functions. The non-linear resistance $\mathcal{R}$ depends on V, t and $\sigma$ that varies over time as described in Eq. (4). Equations (3) and (4) impose a strict framework for resistive switching devices to truly behave as memristive systems. One of the first implications of this definition is that the resistance should vary continuously with V.

In the case of FTJs, we propose that the volume fraction of down domains can be used as the state variable, *i.e.* we identify s with $\sigma$. Then, in the simple case of switching *via* a single zone (a situation that might be achievable in fully patterned, fully epitaxial FTJs), we can write

$$ds/dt = (1-s) \times \left\{ \frac{2}{\tau_P(V)} \left( \frac{t - \tau_N(V)}{\tau_P(V)} \right) \right\} = f(s, V, t) \qquad (5)$$

In contrast with the situation for most other existing memristive systems, we thus reach a description of ferroelectric memristors that goes beyond basic phenomenology and we provide the expression of the function *f* in Eq. (4) for the temporal evolution of the state parameter based on physical arguments.

In summary we have reported transport measurements in ferroelectric tunnel junctions as a function of the amplitude, duration and number of voltage pulses. The resistance can be continuously and reversibly tuned over more than two orders of magnitude by varying the pulse amplitude and/or the pulse number (and thus the total integrated excitation time). These features qualify FTJs as memristive devices. This improves upon previous memristors with a purely electronic mechanism where the resistance contrast is no better than a factor of two[6,7]. Relying on the correlation between junction resistance and ferroelectric domain structure (as imaged by PFM), we model the resistive switching behaviour using a simple model of domain nucleation and growth in a heterogeneous medium. We derive an analytical expression ruling the memristive response, which exemplifies the advantage of resorting to well-established physical phenomena like ferroelectricity in the design of novel memristive systems. Our results invite additional investigations of switching dynamics in nanoscale ferroelectrics and open unforeseen perspectives for ferroelectrics in next-generation neuromorphic computational architectures.

# Acknowledgements

Financial support from the European Research Council (ERC Advanced Grant No. 267579 and ERC Starting Grant No. 259068) and French Agence Nationale de la Recherche (ANR) MHANN is acknowledged. X.M. acknowledges Herchel Smith Fellowship support.

# Figure legends

**Figure 1 : Tuning resistance and ferroelectric domain configuration with voltage amplitude.** (a) Dependence of the junction resistance measured at $V_{read}$=100 mV after the application of 20 ns voltage pulses ($V_{write}$) of different amplitudes. The different curves correspond to different consecutive measurements, with varying maximum (positive or negative) $V_{write}$. (b) Variation of a similar capacitor resistance with the relative fraction of down domains extracted from the PFM phase images. Red-(blue-)framed images show states achieved by the application of positive (negative) voltage pulses of increasing amplitude starting from the ON (OFF) state. The blue and red symbols correspond to the experimental resistance value as a function of the fraction of down domains extracted from the PFM phase images; the black curve is a simulation in a parallel resistance model. The sketch in the top left illustrates the conductance in parallel through the up (low resistance) and down (high resistance) domains.

**Figure 2 : Pulse duration / pulse number phase diagrams.** Resistance of a junction for different pulse duration repeated $N_{write}$ times for three pulse amplitudes.

**Figure 3 : Tuning resistance by consecutive identical pulses.** (a, c) Evolution of the junction resistance as a function of the different voltage pulse sequences [plotted in (b) for $V_{write}$ = +2.9 V and –2.7 V and in (d) for $V_{write}$=+3 V and -3 V].

**Figure 4 : Polarization switching dynamics.** (a-f) Dependence on the switched fraction with the cumulative pulse time for down-to-up (a-c) and up-to-down (d-f) switching and different voltage amplitudes. The data are shown as symbols and the lines are fits (see text for details). Evolution of the nucleation and propagation times with the inverse of the applied electric field, for negative (g-h) and positive (i-j) bias. The symbol size is proportional to the corresponding fraction of switched area.

# Methods

*Samples :*

The BaTiO$_3$/La$_{0.67}$Sr$_{0.33}$MnO$_3$ (BTO/LSMO) bilayers were grown on (001)NdGaO$_3$ single-crystal substrates by pulsed laser deposition [KrF excimer laser (λ=248 nm), fluence of 2 J/cm², repetition rate of 1Hz]. 30-nm LSMO films were grown at 775°C under 0.15 mbar of oxygen pressure. BTO films were subsequently grown at 775°C and 0.10 mbar oxygen pressure. The samples were annealed for 1 hour at 750°C and 500 mbar oxygen pressure and cooled down to room temperature at 10°C/min. The thickness of the films was calibrated with X-ray reflectivity and crosschecked with transmission electron microscopy. The nanodevices with diameters of 350 nm were defined from these bilayers by electron-beam lithography and lift-off of sputter-deposited Co (10 nm) followed by a capping layer of Au (10 nm).

*Measurements*

Electrical measurements were performed with a Digital Instruments Nanoscope IV set-up at room temperature and under nitrogen flow with commercial Si tips coated with Cr/Pt (Budget Sensors). The bias voltage was applied to the tip and the sample was grounded for electrical measurements. For voltage pulses time widths below 500 ns, a bias tee was connected to the AFM to split voltages pulses from DC measurements. An Agilent 81150A pulse generator was used to apply voltage pulses of duration of 10 ns to 200 ns and resistances after the applied pulses were measured with a Keithley 6487 picoammeter using a Yokogawa GS610 voltage source at 100 mV.

Piezoresponse force microscopy (PFM) experiments were performed with a multimode Nanoscope IV set-up and SR830 lock-in detection. A TT*i* TG1010 external source was used to apply a 12kHz ac sinusoidal excitation of 1V peak to peak with a dc offset of 100mV. The tip was grounded for PFM experiments. Successive PFM images were collected after setting the device to a chosen resistance state by application of 100 µs voltage pulses.

*Switching dynamics model*

The model is based on the following assumptions :

- the switching occurs through zones with different parameters in terms of domain wall propagation speed, nucleation time, number of nuclei
- each zone follows the KAI model
- for a given zone starts, we suppose that all nucleation sites are activated at the same time t = $\tau_N$, the nucleation time. $\tau_N$ depends on the voltage.
- the number of zones involved in the switching process depends on the voltage.

Following this set of assumptions, the ratio s can be written as :

$$s = \sum_{i=1}^{N(i)} S_i * h(t - \tau_N^i) * \left\{ 1 - \exp\left[ -\left( \frac{t - \tau_N^i}{\tau_P^i} \right)^2 \right] \right\}$$

For down to up switching, and

$$s = 1 - \sum_{i=1}^{N(i)} S_i * h(t - \tau_N^i) * \left\{ 1 - \exp\left[ -\left( \frac{t - \tau_N^i}{\tau_P^i} \right)^2 \right] \right\}$$

With : $\sum_{i=1}^{N(i)} S_i = 1$

$S_i$ is the area of each zone normalized by the total junction area. h is the Heaviside step function.

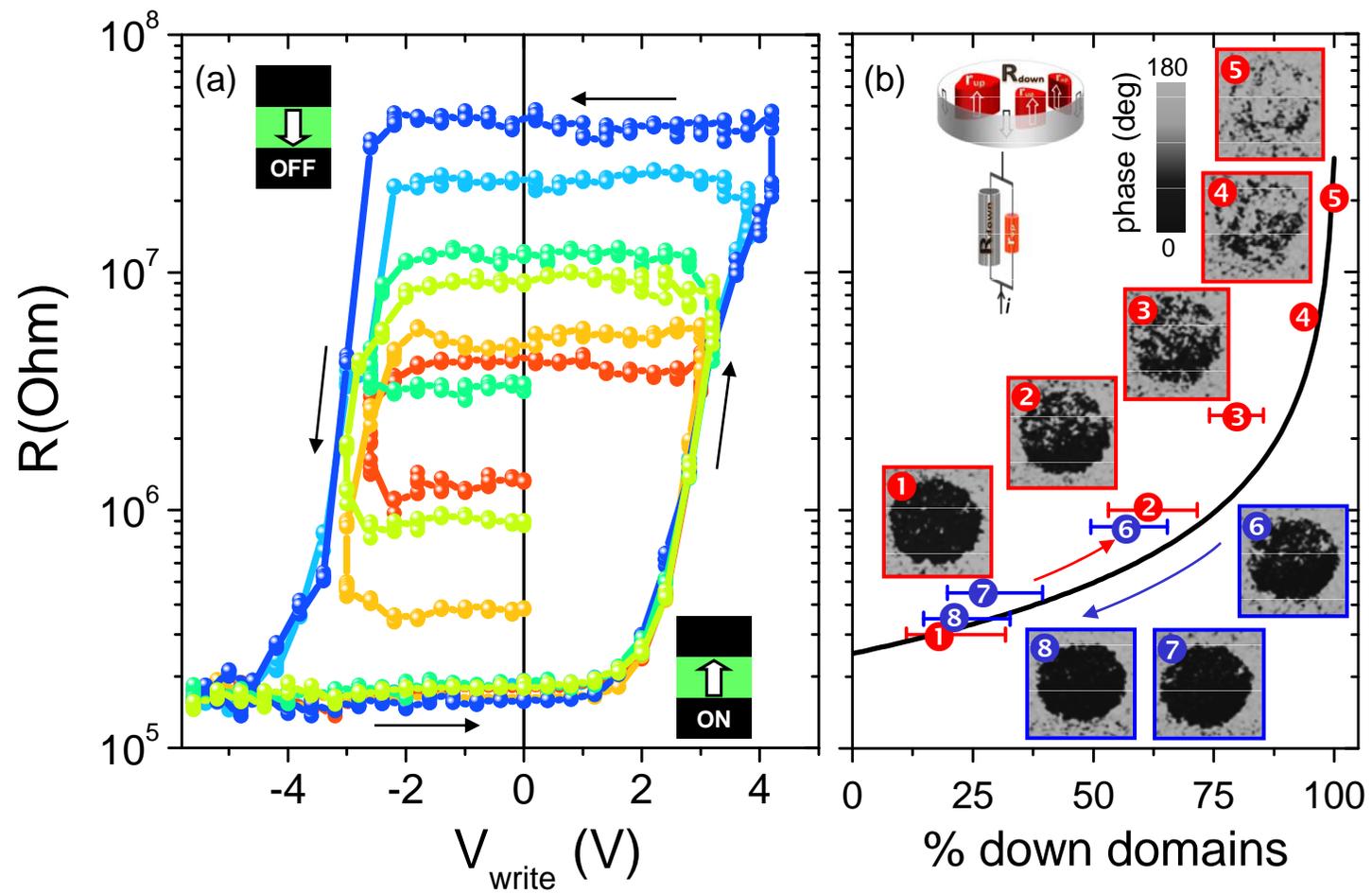

Fig. 1 : Chanthbouala et al

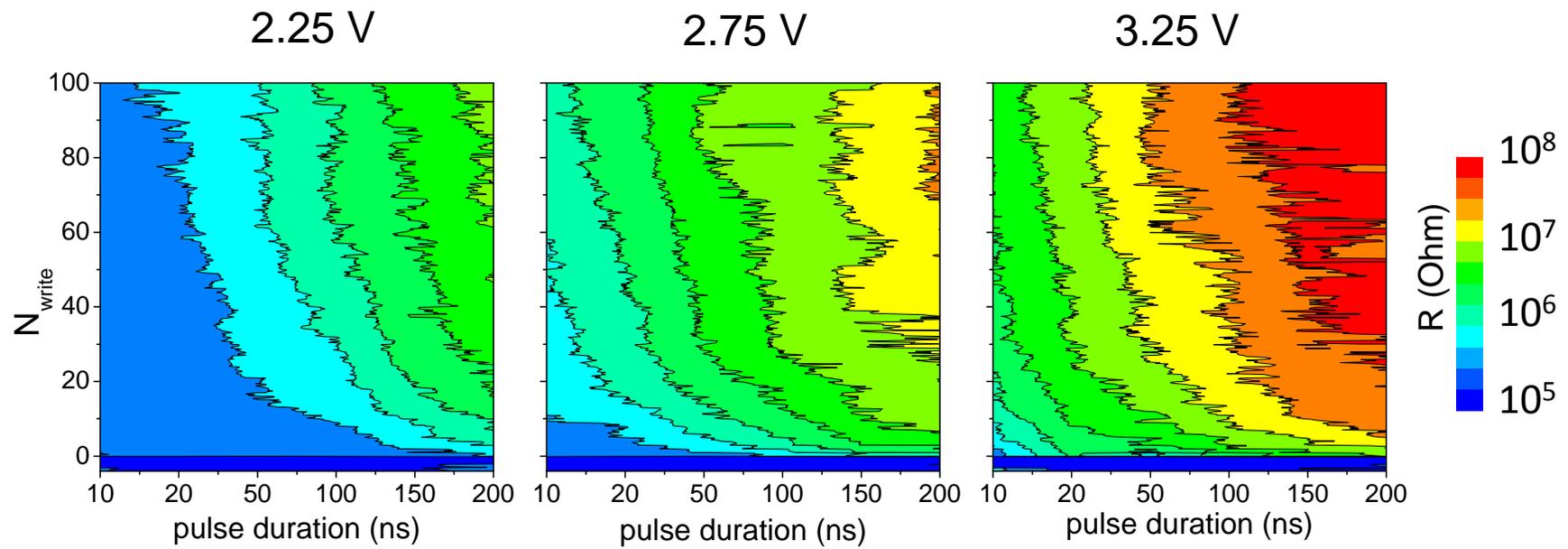

Fig. 2 : Chanthbouala et al

Fig. 3 : Chanthbouala et al

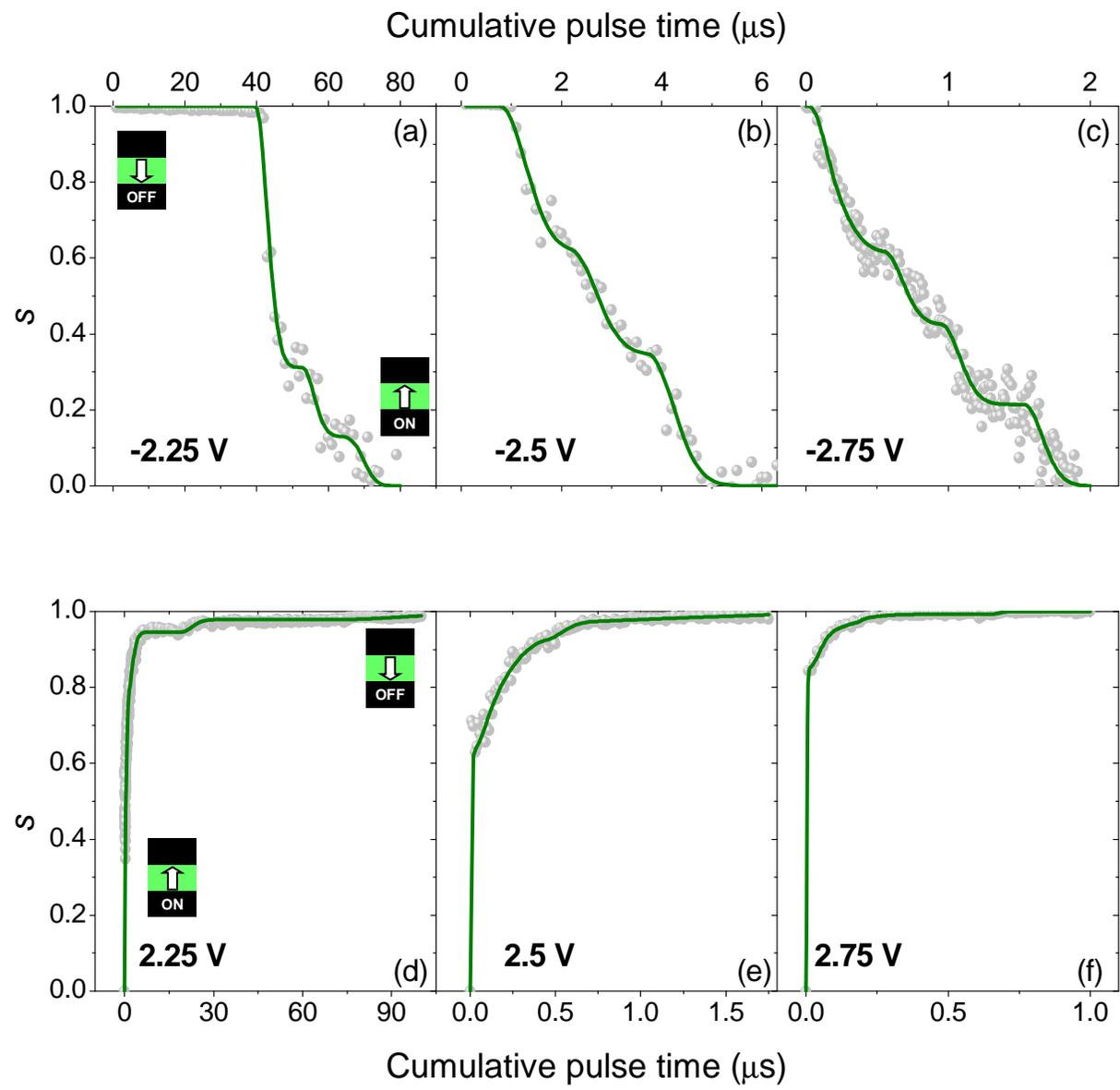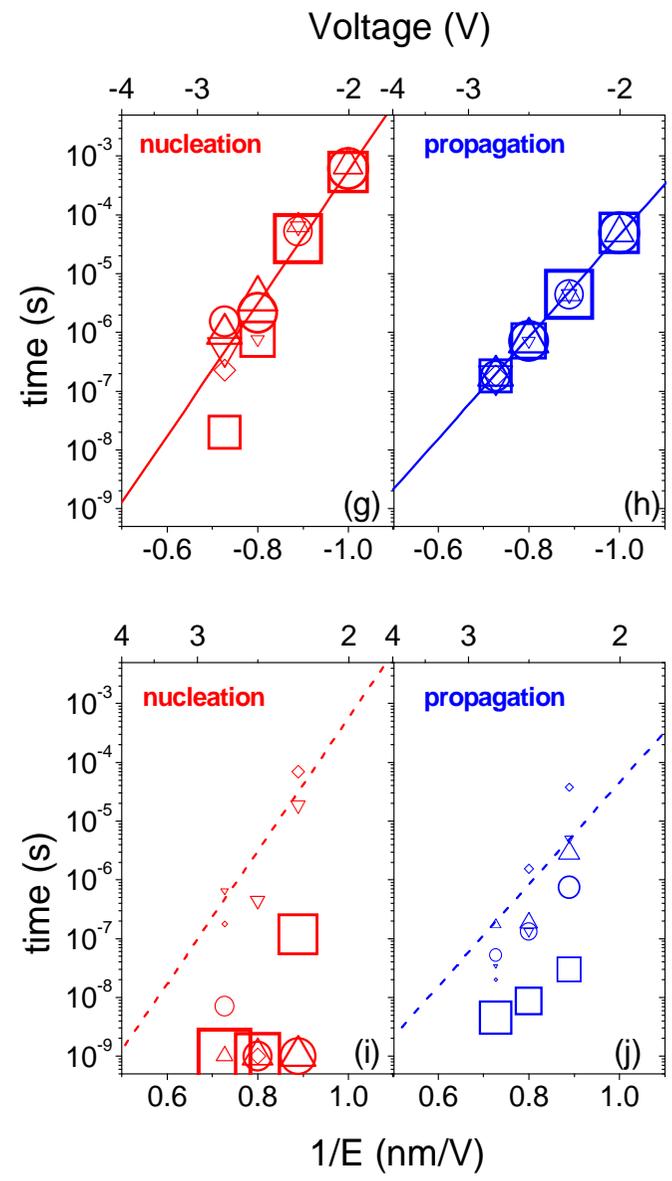

Fig. 4 : Chanthbouala et al